\documentclass{aa}  
\usepackage{graphicx}
\usepackage{txfonts}
\usepackage{hyperref}
\graphicspath{{./}{figures/}}
\usepackage[normalem]{ulem}
\begin{document}

   \title{LOFAR observations of radio burst source sizes and scattering in the solar corona}

   \author{Pearse C. Murphy
          \inst{1} \inst{2}
          \and
          Eoin P. Carley
          \inst{2}
          \and
          Aoife Maria Ryan
          \inst{1} \inst{2}
          \and
          Pietro Zucca
          \inst{3}
          \and
          Peter T. Gallagher
          \inst{2}
          }

   \institute{School of Physics, Trinity College Dublin, Dublin 2, Ireland.
              \email{pearse.murphy@dias.ie}
         \and
            Astronomy \& Astrophysics Section, Dublin Institute for Advanced Studies, D02 XF86, Ireland.
        \and
            ASTRON,   Netherlands   Institute   for   Radio   Astronomy, Postbus 2, 7990 AA, Dwingeloo, The Netherlands.
             }

   \date{Received May 2020; accepted;}

\abstract{
Low frequency radio wave scattering and refraction can have a dramatic effect on the observed size and position of radio sources in the solar corona.
The scattering and refraction is thought to be due to fluctuations in electron density caused by turbulence. Hence, determining the true radio source size can provide information on the turbulence in coronal plasma.
However, the lack of high spatial resolution radio interferometric observations at low frequencies, such as with the LOw Frequency ARray (LOFAR), has made it difficult to determine the true radio source size and level of radio wave scattering.
Here we directly fit the visibilities of a LOFAR observation of a Type IIIb radio burst with an elliptical Gaussian to determine its source size and position. This circumvents the need to image the source and then de-convolve LOFAR's point spread function, which can introduce spurious effects to the source size and shape.
For a burst at 34.76~MHz, we find full width at half maximum (FWHM) heights along the major and minor axes to be $18.8^\prime$~$\pm~0.1^\prime$ and $10.2^\prime$~$\pm~0.1^\prime$, respectively, at a plane of sky heliocentric distance of 1.75~R$_\odot$.
Our results suggest that the level of density fluctuations in the solar corona  is  the  main  cause  of  the  scattering  of  radio  waves, resulting in  large  source sizes. However, the magnitude of $\varepsilon$ may be smaller than what has been previously derived in observations of radio wave scattering in tied-array images.}


   \maketitle

\section{Introduction} \label{sec:intro}
Low frequency radio wave propagation in the solar corona is not fully understood. It is widely accepted that the scattering of radio waves off of density inhomogeneities plays a key role in the observed source sizes of radio bursts \citep{Fokker1965,Steinberg1971,Stewart1972,Riddle1974,Thejappa2007,Thejappa2008,Kontar2019}. However, the exact extent to which observed source sizes are broadened is difficult to measure as it requires an angular resolution to spatially resolve the source as well as an \textit{a priori} knowledge of its original size.
Current generation radio interferometers such as the LOw Frequency ARray \citep[LOFAR;][]{VanHaarlem2013b} have the resolving power to observe this angular size. This angular resolution can be exploited to accurately determine burst size and position, both of which are indicators of the level of radio scattering in the corona, which in turn is related to the level of turbulent density fluctuations. Hence, a better understanding of scattering may lead to new insights into the nature of coronal turbulence.

The study of radio wave scattering in the solar corona has its origins in the 1960s and 1970s. \cite{Fokker1965}, \cite{Steinberg1971}, \cite{Stewart1972}, and \cite{Riddle1974} conducted seminal work on the ray tracing of radio waves in various coronal models. They all concluded that sources emitted near the plasma frequency in the solar corona are enlarged due to the scattering of radio waves from coronal density fluctuations. While the explanation of coronal scattering for observed source characteristics fell out of favour by the mid-1980s \citep{McLean1985}, it has seen a renewed interested in low frequency radio observations in recent years  \citep{Thejappa2007,Thejappa2008,Kontar2017,Sharykin2018,Gordovskyy2019,Kontar2019}.

In low frequency imaging, the extent of scattering in the corona can be determined through the analysis of Type III radio bursts, particularly their position and size in images or decay times in dynamic spectra \citep[e.g.][]{Kontar2019, Gordovskyy2019, Krupar2018}. Given that these bursts are due to plasma emission from electron beams propagating through coronal plasma \citep[see][for a review]{Reid2014}, they provide a density diagnostic of such plasma. In particular, a subset of these bursts known as `Type IIIb' provides a diagnostic of scattering in coronal plasma due to density fluctuations from turbulence. For example, Type IIIb bursts often show fine structures or `striae' along the burst envelope \citep{Ellis1967,Ellis1969,DeLaNoe1972,DeLaNoe1975,Melnik2010b}, which are believed to be caused by density inhomogeneities in the corona \citep{Takakura1975}. Using a density model, the frequency bandwidth of these striae can be used to infer the vertical extent of the density inhomogeneity in space. A comparison of this spatial extent to observed source sizes in images can provide the extent to which the radio emission has been scattered \citep[e.g.][]{Kontar2017}.

Theoretically, the extent of scattering in the corona is related to the root mean square (r.m.s) fluctuations of electron density $\varepsilon = \sqrt{\left< \delta n^2 \right>}/n$. Many recent works have assumed a value for $\varepsilon$ to use in simulations in order to recreate the time profile and source size of solar radio bursts \citep[e.g.][]{Krupar2018, Kontar2019}. However, few have used the observed source size and time profile to determine $\varepsilon$. Those that have are limited to determining the value of $\varepsilon$ in the solar wind at distances $> 10 \, R_\odot$. Techniques such as interplanetary scintillations \citep[e.g.][]{Bisoi2014} and crab nebula occultation \citep{SasikumarRaja2016} have also been used to determine $\varepsilon$ at these distances. The general conclusion of these studies is that $\varepsilon$ varies slowly with heliocentric distance and has typical values of $0.001 \lesssim \varepsilon \lesssim 0.02$ in the range of 10 to 45 $\, R_\odot$. Despite this, larger values of $\varepsilon$ have been used in models. For example, \cite{Reid2010} used a value of $\varepsilon \approx 0.1$ to model electron beam transport, while \cite{Kontar2019} recently used Monte Carlo simulations of scattering to determine that a value of $\varepsilon = 0.8$ is necessary in order to account for source sizes of the order of $20^\prime$, as observed by \cite{Kontar2017}. Measured values of $\varepsilon$, particularly at heights of ${\sim} 2 \, R_\odot$, are not common in the literature, with the exception of a recent study by \cite{Krupar2020}. By using observations from the Parker Solar Probe \citep[PSP,][]{Fox2016}, \cite{Krupar2020} calculate a value for $\varepsilon = 0.07$ at a plasma frequency $f_p = 137$ kHz. They also find that the value of $\varepsilon$ decreases from 0.22 to 0.09 over a height range of 2.4 to 14$\, R_\odot$.

It is clear that a correct interpretation of radio observations provides a means to investigate the level of scattering and density fluctuation in the corona. Advances in radio astronomy over the past 40 years have led to increased sensitivity, temporal resolution, frequency resolution, and resolving power. Modern radio telescopes such as LOFAR, the Murchison Widefield Array \citep[MWA;][]{Lonsdale2009}, and the upcoming Square Kilometre Array \citep[SKA;][]{Dewdney2009} are capable of observing the predicted spatial and time profiles of Type IIIb bursts. That said, previous studies with LOFAR have tended to use tied-array imaging in this regard \citep{Kontar2017}, which has limited spatial resolution with respect to interferometric imaging.

In this paper, we use LOFAR interferometric observations to determine the observed radio source size and position and how this differs from the expected source properties, which can be estimated from spectroscopy. %The discrepancy between what is observed and what is expected is used to determine the relative level of density fluctuations in the corona. 
Directly fitting interferometric visibilities provides us with the opportunity to observe low frequency radio sources at a spatial resolution in excess of what has usually been achieved. We compare our results with those of the tied-array observation from \cite{Kontar2017} and discuss the implication this may have on determining the relative level of density fluctuations in the corona. 
The remainder of this paper is outlined as follows: An observation of a Type IIIb burst is described in Sect. \ref{sec:obs}. In Sect. \ref{sec:data}, we detail a method of directly fitting interferometric visibilities in order to recreate a sky-brightness distribution and give results of observed source sizes. Section \ref{sec:data} also includes an analysis of a Type IIIb stria. We conclude with a discussion in Sect. \ref{sec:conclusion}.

\section{Observation} \label{sec:obs}
LOFAR is an interferometric array that spans across Europe, observing radio frequencies at 10 - 240 MHz. 
An interferometric observation of the Sun, utilising 36 stations (24 core and 12 remote), was performed on 17 October 2015 from 08:00 UTC to 14:00 UTC. During this time, a Type III solar radio burst was recorded  at 13:21 UTC. A calibrator source, Virgo A, was observed co-temporally in all sub-bands over the course of the observation.

\begin{figure}
    \centering
    \includegraphics[width=\columnwidth]{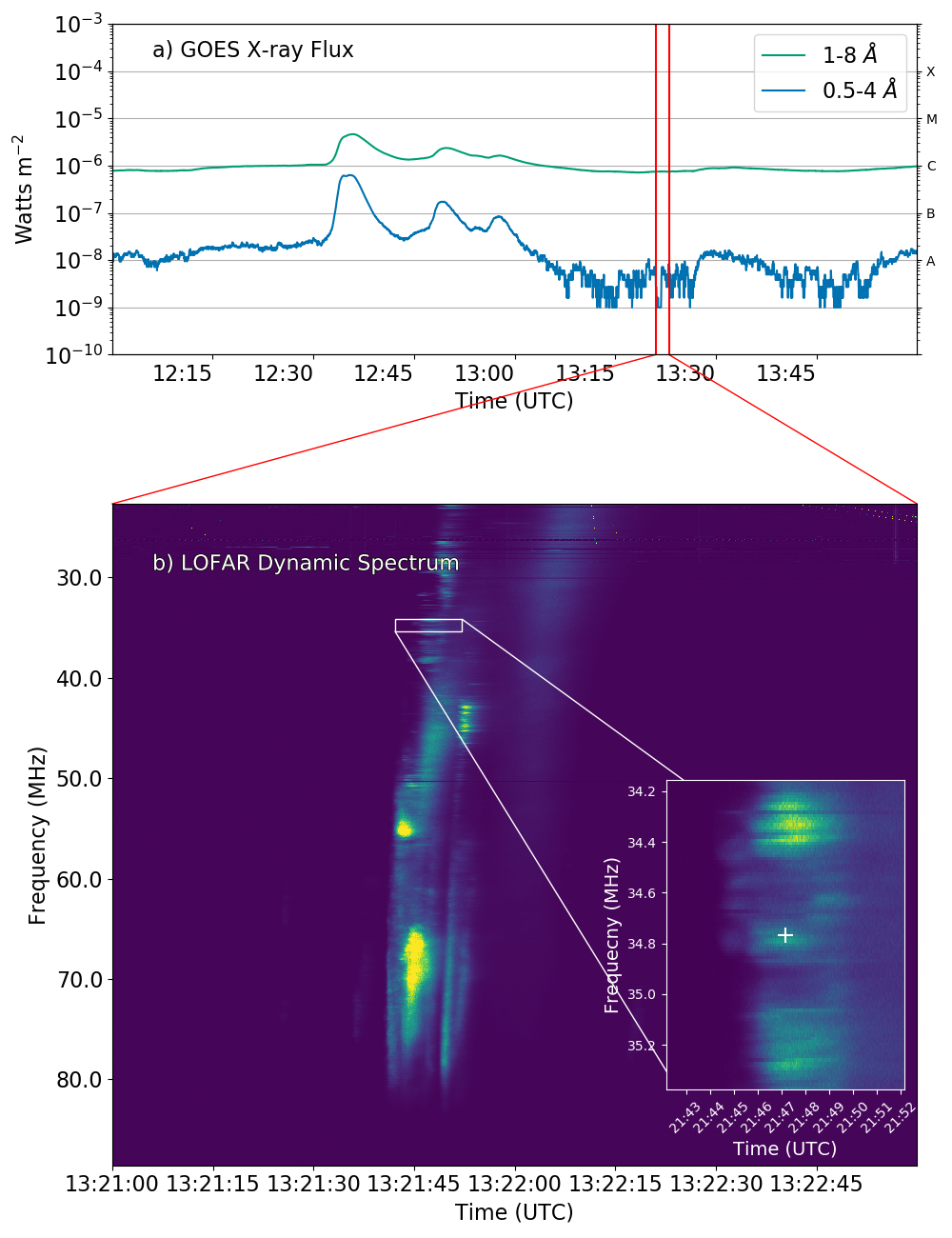}
    \caption{Overview of the X-ray flux and LOFAR dynamic spectrum for the observation on 17 October 2015. \textbf{a)} GOES X-ray lightcurves for the duration of the LOFAR solar observation. Minimal activity other than a number of C-class flares prior to 13:00 UTC is observed. Red vertical lines indicate the time range of the radio analysis. \textbf{b)} Dynamic spectrum of a Type III solar radio burst observed with LOFAR station RS509. We note the striations in frequency that  are particularly apparent below 40 MHz. The inset is a zoom-in of the region in the white box, which shows striation in the burst. The white cross indicates the time and frequency at which the images described in Sect. \ref{sec:data} are made.}
    \label{fig:context}
\end{figure}

Figure \ref{fig:context}a shows the X-ray flux measured by the Geostationary Operational Environmental Satellite (GOES) for the duration of the LOFAR observation. A number of C-class flares can be seen in in Fig. \ref{fig:context}a, but no significant activity is noticeable at the time of the radio burst, which is indicated by the red vertical lines

A dynamic spectrum of the burst was recorded in the Low Band Antenna (LBA) band by remote station RS509 and is shown in Fig. \ref{fig:context}b. The inset shows a number of striations from 34 - 35 MHz, and the white cross indicates the time and frequency at which the images described in Sect. \ref{sec:data} are made.

The maximum baseline of the LOFAR observation is 84\,km.\ This provides sub-arcminute resolution across almost all of the observed frequency range, offering an unprecedented level of spatial resolution.

\section{Data analysis and results} \label{sec:data}
The source sizes and positions of solar radio bursts in LOFAR data have typically been obtained by the `tied-array imaging mode' \citep{Morosan2014} whereby a number of beams are tessellated across the Sun and the response in each beam is interpolated to produce an image \citep[e.g.][]{Reid2017,Kontar2017,Zucca2018, Morosan2019b}. Tied-array imaging has the distinct advantage over interferometric observations in that it retains the ${\sim} 12$ kHz frequency resolution and ${\sim} 0.01$ s temporal resolution from LOFAR beam-formed observations, but it also contains a significant limitation. Tied-array observations can only be made using the LOFAR core stations as they share a single clock \citep{DeGasperin2019}, which makes it possible to add beam-formed data coherently. This means that the maximum baseline from tied-array observations is approximately 2 km, which corresponds to an angular resolution of ${\sim} 17^\prime$ at ${\sim} 30$\,MHz. Furthermore, the effect of interpolation between each tied-array beam on the observed source size has not yet been compared to observations done interferometrically. It is therefore unclear whether previously observed source sizes are in fact due to the underlying source or if they are an effect of the imaging technique. Solar campaigns with LOFAR are now performed with a new mode that allows for simultaneous interferometric and tied-array observations. A detailed comparison of these modes is currently being studied, which should resolve the ambiguity in source sizes determined with tied-array observations (Morosan, D. E. 2020, private communication). 
%a result of this interpolation or

In order to avoid such limitations of the tied-array mode, here we use interferometric observations from the LOFAR core and remote stations, which offer a longer baseline of 84\,km and hence much better spatial resolution.
The LOFAR data from this observation were calibrated using the Default Preprocessing Pipeline \cite[DPPP;][]{VanDiepen2018} and a co-temporal observation of Virgo A. This corrects for effects such as antenna band-pass, clock drift, and propagation effects through the ionosphere \citep{DeGasperin2019}. We next describe our technique of directly fitting the LOFAR visibilities to estimate radio source size and position.

To produce an image from interferometric observations, an inverse Fourier transform was performed on the observed visibilities, usually followed by a de-convolution of the array point spread function (PSF) from the resulting `dirty-map' of the sky-brightness distribution.
For such a de-convolution, LOFAR uses an implementation of the multi-scale CLEAN algorithm known as WSClean \citep{Offringa2014}. In this procedure, a weighting may be applied to the visibilities to improve the sensitivity to various spatial scales on the radio sky, the most common of which is the Briggs robustness weighting scheme \citep{Briggs1995}. Recreating a radio image in this way can introduce artefacts, depending on the Briggs robustness used, the number of iterations of the algorithm, and a number of other parameters described in more detail in, for example, \cite{Hogbom1974, Cornwell2008, Offringa2014, Offringa2017}. These artefacts include changes to the source shape and size. Therefore, to avoid ambiguity in the source size, shape, and position due to such imaging algorithms, we directly fitted the measured visibilities, similar to a method used for X-ray observations using the Reuven Ramaty High Energy Solar Spectroscopic Imager \citep[RHESSI][]{Hurford2002, Kontar2010}. We describe our method in the following subsections. 

\subsection{Fitting the visibilities}
\label{sec:vis_fit}
The \textit{uv} plane is a Fourier space representation of antenna pair positions. Each point in the \textit{uv} plane is sensitive to the emission of a particular angular scale. Due to the timescales over which Type III bursts occur, solar observations are limited to a sparse sample of the \textit{uv} plane, and techniques to increase samples in this plane, such as aperture synthesis, cannot be used. However, the large brightness temperatures of Type III and Type IIIb radio bursts \citep{Reid2014} give rise to a  high signal-to-noise ratio, which allows a direct fit of a model to the visibilities.
In the following, we assume that the emitting source is a single elliptical Gaussian. This is based on the dynamic spectrum in Figure \ref{fig:context}b, which shows that the Type IIIb burst does not overlap any other bursts and as such is probably the only source in an interferometric image. This assumption leads to the convenient fact that an elliptical Gaussian in real space is observed as another elliptical Gaussian in the \textit{uv} plane. The form of this Gaussian is
\begin{equation}
V(u,v) = e^{-2\pi i(ux_0+vy_0)} \left( \frac{I_0}{2\pi} e^{-\left(\frac{\sigma_x^2(2\pi u^\prime)^2}{2}-\frac{\sigma_y^2(2\pi v^\prime)^2}{2}\right)} + C \right)
\label{eq:vis_gauss}
,\end{equation}
where $x_0$ and $y_0$ are the x and y coordinates of the source centre in real space, $\sigma_x$ and $\sigma_y$ are the standard deviation in the x and y directions, and $C$ is a constant background. Here the visibilities have been rotated to a new coordinate frame with axes $u^\prime$ and $v^\prime$. These axes are parallel and perpendicular to the major and minor axes of the Gaussian source such that $u^\prime = u\cos{\theta} - v\sin{\theta} \mbox{ and } v^\prime = u\sin{\theta} + v\cos{\theta}$,  where $\theta$ is the angle of the major axis to the x axis (i.e. the position angle of the Gaussian on the \textit{uv} plane). 

A non-linear least squares fit was applied to the sample of the visibilities in two stages. First, the source size, maximum intensity, and angle relative to the x axis were found by fitting the absolute value of the complex visibilities.
In order to determine the source location, the phase angle of the data was fitted. Source location in real space determines the fringe separation and orientation in Fourier space. The direct fitting of parameters to \textbf{$V(u,v)$} was then used to recreate the sky-brightness distribution or image \textbf{$I(x,y)$}, which is the inverse Fourier transform of Eq. (\ref{eq:vis_gauss}).

\begin{figure}
    \centering
    \includegraphics[width=\columnwidth]{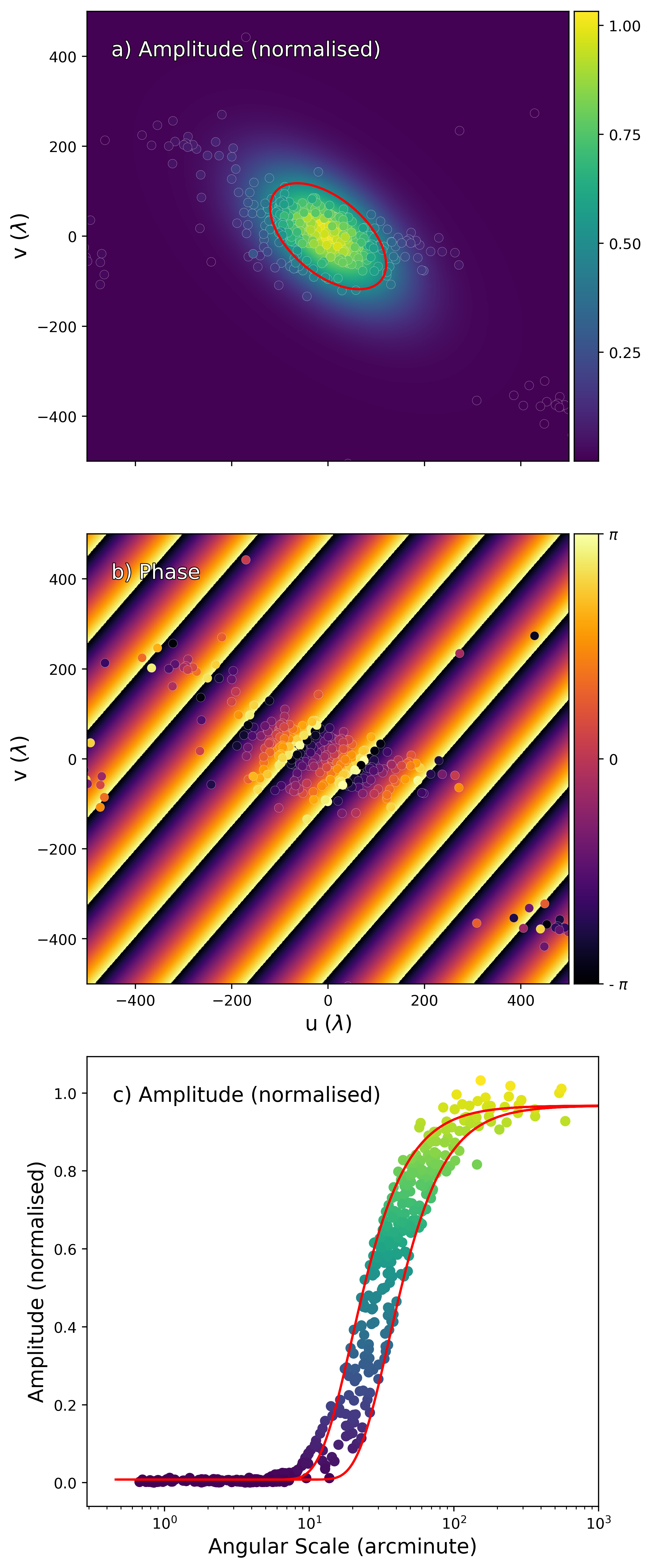}
    \caption{Results of directly fitting the LOFAR visibilities. \textbf{a)} Amplitudes of visibilities in the \textit{uv} plane for the LOFAR observation. The background colour map shows a Gaussian fit. The red ellipse shows the FWHM of the normalised amplitude. \textbf{b)} Visibility phase in the \textit{uv} plane. The background colour map shows the fitted phase angle. \textbf{c)} Visibility amplitudes received from different angular scales. The red curves indicate the FWHM of the semi-major and semi-minor axes of the fitted Gaussian.}
    \label{fig:uv_fit}
\end{figure}

Figure \ref{fig:uv_fit} shows the fit of the modelled Gaussian to the complex visibilities. Due to the fact that this fit is done in Fourier space, the amplitude and phase of the data and fit are shown in the \textit{uv} plane in Figs. \ref{fig:uv_fit}a and \ref{fig:uv_fit}b, respectively.  Here, the points are the observed visibilities and the background colour map is the fit. In Fig. \ref{fig:uv_fit}a, a red ellipse indicates the full width at half maximum (FWHM) height of the fitted Gaussian. The fringes in Fig.~\ref{fig:uv_fit}b show the fit of the source position to the distribution of visibility phases across \textit{uv} space.
Figure \ref{fig:uv_fit}c shows the increase in the amplitude of recorded visibilities with the angular scale on the sky that causes this increase. The red curves are where data points would lie for a Gaussian in visibility space, with the FWHM in the major and minor directions obtained from the fit in Fig. \ref{fig:uv_fit}a.

\begin{figure}
    \centering
    \includegraphics[width=\columnwidth]{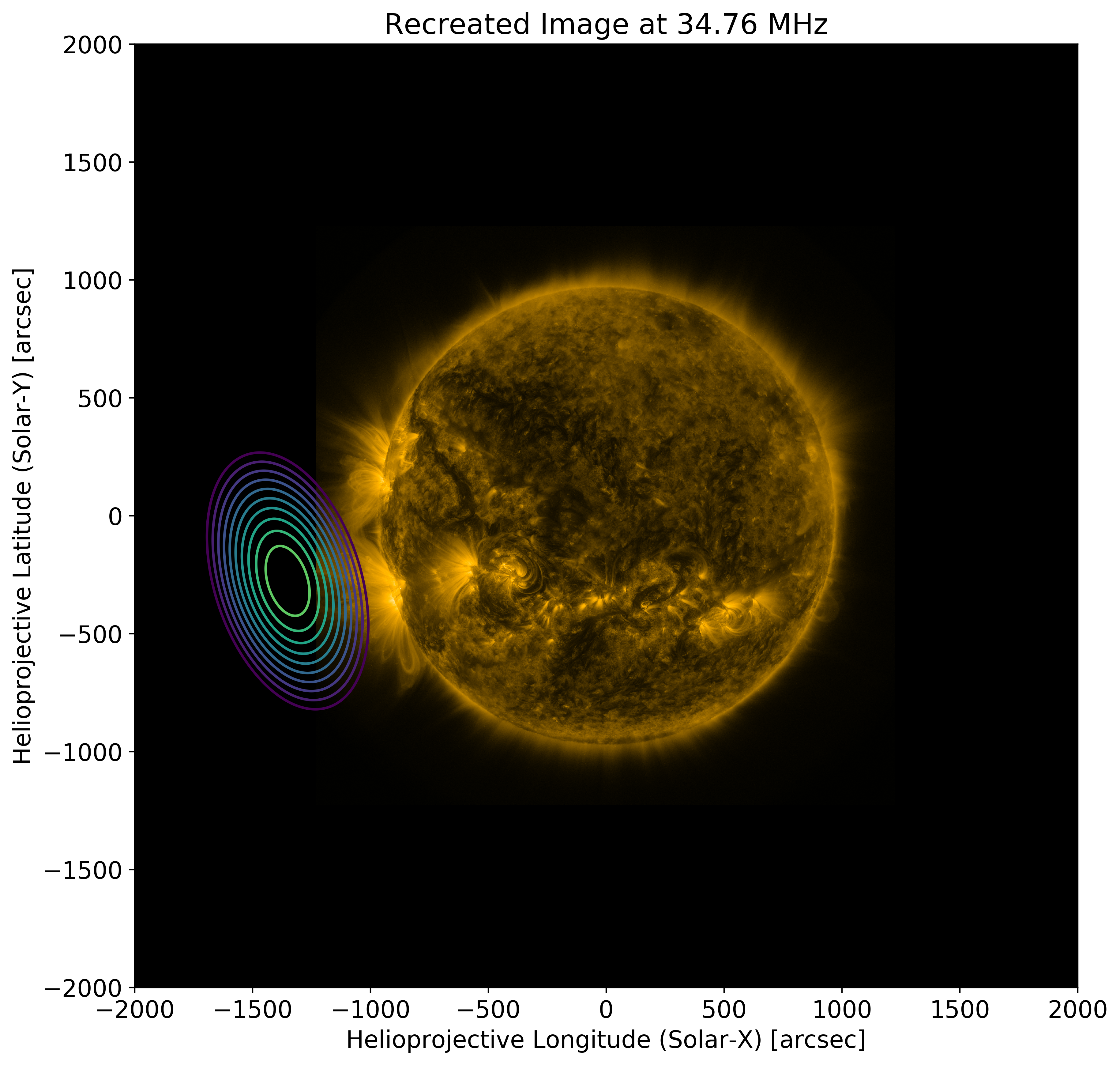}
    \caption{Recreated sky intensity profile of a Type IIIb radio burst occurring at 13:21:46 on 17 October 2015. The background solar image is an AIA 171$\AA$ image from 17 October 2015 at 13:21:46 UTC.} 
    \label{fig:recreate}
\end{figure}

The visibility fit reveals a source with FWHMs in real space of $18.8^\prime$ $\pm 0.1^\prime$ and $10.2^\prime$ $\pm 0.1^\prime$, in the direction of the major and minor axes, respectively. The source is found at a position of $-1312^{\prime\prime}, -1064^{\prime\prime}$ from the solar centre, giving a plane of sky distance of 1.75$\, R_\odot$. The parameters from the fit can then be used to recreate a sky-brightness distribution $I(x,y)$ in real space, which is shown as contours over-plotted on a 171\,\AA~ image taken by the Atmospheric Imaging Assembly \citep[AIA;][]{Lemen2012} in Fig. \ref{fig:recreate}. We note that despite the theoretical high angular resolution of the long baselines afforded by LOFAR remote stations (84\,km), the source size is still large and we see little evidence of angular scales smaller than $\sim$10$^\prime$ in Fig.~\ref{fig:uv_fit}c. We will discuss this further in Sect. \ref{sec:conclusion}.

\subsection{Type IIIb striae}
In the above section we determined the source size and position using a direct modelling of LOFAR visibility observations. This provides us with an opportunity to compare the observed source size to its actual size, which can be estimated from spectroscopic observations, similar to the method in \cite{Kontar2017}.

To estimate the source size from spectroscopic measurements, we related the FWHM of the frequency of the striation to its vertical extent in the solar corona $\Delta r {\sim} 2L \left(\Delta f/f\right)$, where $L$ is the characteristic density scale height \citep{Kontar2017}.
A single striation was manually identified at 34.76 MHz from the dynamic spectrum. The time of maximum intensity for the burst was found, and a vertical frequency slice was obtained, from which $\Delta f/f$ was calculated. The individual striation, the centre of which is indicated by a white cross in the inset panel of Fig. \ref{fig:context}b, was fit with a Gaussian. The value for $\Delta f$ of the striation was found from the FWHM of the fitted peak to be $\Delta f \sim 0.2$~MHz. The frequency-to-bandwidth ratio for the striation was found to be $\Delta f/f = 0.006$, leading to an estimated source size of $3.18^{\prime\prime}$. 
Similar to \cite{Kontar2017}, this is far smaller than the source size observed from the visibility fit.

In the following section we will discuss why the most probable cause of the discrepancy in source size is radio scattering. We will also provide a discussion on the comparison of this observation to recent developments in the theory as well as the effect that scattering has had on actual source size and position.

\section{Discussion}\label{sec:conclusion}

Adopting the theory described by \cite{Takakura1975} and used in \cite{Kontar2017}, the predicted source sizes of a Type IIIb striation are much smaller than what is observed. The most probable cause of this discrepancy is a combination of radio light scattering in the solar corona, propagation effects in the Earth's ionosphere, and limitations due to angular resolution.
With this observation, we accounted for and corrected ionospheric effects in the calibration step \citep[Sect. \ref{sec:data} and ][]{DeGasperin2019}, thereby removing the largest uncertainty in source size and position. By fitting the source size directly in visibility space, we can directly see the power at which different angular scales were observed. The \textit{uv} coverage of this observation allows angular scales of ${\sim} 42^{\prime\prime}$ to be observed (Fig. \ref{fig:uv_fit}c); although this is larger than the predicted source size of ${\sim}3^{\prime\prime}$, the amplitude of the observed visibilities does not increase until ${\sim} 10^\prime$, indicating that it is, in fact, the smallest source size observed in the visibilities.

It should be noted that there is better \textit{uv} coverage along one axis compared to its orthogonal, which may have an effect on the eccentricity of the elliptical Gaussian fit. However, owing to the qualitatively similar shape as predicted by \cite{Kontar2019} for a burst originating near the solar limb, we are confident that the eccentricity is representative of the real source. The orientation and elongation of the source are also consistent with observations of anisotropic scattering in the solar wind \citep{Anantharamaiah1994, Ingale2015}, where scattered sources are elongated perpendicular to the large-scale (radial) magnetic field of the Sun.

Having accounted for all systematic effects that can affect the source size, we conclude that the large source sizes observed in this observation are due to the effect of scattering in the solar corona only. Previous tied-array observations have led to similar conclusions; however, the tied-array technique interpolates data from a tessellation of beams across the Sun, and this introduces an ambiguity to the origin and size of sources. 
We assume the origin of the radio source to be somewhere above the active regions close to the east limb. 
An exploratory potential free source surface (PFSS) extrapolation suggests open field lines at an angle of $ \theta_s {\sim} 20^\circ$ from the plane of sky towards the observer. Similar to \cite{Chrysaphi2018} (Eqs. 5 and 6), we determined an out-of-plane heliocentric distance for the source. Using the observed in-plane heliocentric distance and an angle of $ \theta_s {\sim} 20^\circ$ from the plane of sky, we obtain an out-of-plane heliocentric distance of 1.82$\, R_\odot$. Comparing this to Fig. 8 in \cite{Kontar2019}, which shows the effect of the angle from the plane of sky $\theta_s$ on source position and FWHM size in the major and minor axes, we would expect a ratio of the FWHM on the minor axis to the FWHM of the major axis to be of the order of 0.6. Our observations show a ratio of 0.54, suggesting $ \theta_s {\sim} 20^\circ$ is an appropriate approximation for the angle from the plane of sky.

The FWHMs of the source at 34.76~MHz along the major and minor axes for this observation are $18.8^\prime$ $\pm 0.1^\prime$ and $10.2^\prime$ $\pm 0.1^\prime$, respectively. This gives the FWHM area of the source to be $A_s = 150.6~\mbox{arcmin}^2$, which we note is smaller than that of \cite{Kontar2017}, who measure $A_s = 400~\mbox{arcmin}^2$ at a similar frequency of 32.5~MHz. While this could simply be due to these being two separate observations, it may be more indicative of a discrepancy between source sizes measured in interferometric observations and tied-array observations.
As mentioned in Sect.~\ref{sec:data}, the spatial resolution of LOFAR interferometric observations is superior to that of tied-array observations. This is mostly due to the additional stations that can be used for interferometric imaging and thus greater baseline lengths, but it is also due to the way tied-array images are made. Tied-array observations are carried out by pointing a number of beams in a honeycomb-like pattern centred on the Sun and interpolating data from each of the tessellated beams. The effect of this on observed source sizes and position has not yet been characterised. \cite{Kontar2017} attribute the large source size observed in their tied-array observation of a Type IIIb radio burst to the scattering of radio waves off of density inhomogeneities in the solar corona. It was later determined that a relative r.m.s fluctuation of electron density of $\varepsilon = 0.8$ was necessary to explain the large source size observed by \cite{Kontar2019}. While it is evident that radio wave scattering causes radio bursts to appear larger in observations than predicted, the reduced spatial resolution of tied-array imaging may result in an overestimate of $\varepsilon$.

The last decade has seen a renewed interest in low frequency observations of the radio Sun with the advent of state-of-the-art radio interferometers such as LOFAR and the MWA. Radio bursts emitted via the plasma emission process give a diagnostic of the local plasma density, which may give insight into the turbulent nature of coronal plasma. It is theorised that the size of low frequency radio emission is limited by the scattering caused by turbulence \citep{Bastian1994}; however, the angular resolution necessary to challenge this theory has only recently become available. In particular, a robust comparison of sources observed with tied-array and interferometric imaging is needed. Analytical approximations of radio scattering \citep[e.g.][]{Chrysaphi2018,Gordovskyy2019,Sharma2020} have had some success in accounting for the apparent source shift and brightness temperature due to scattering; however, they cannot account for the anisotropic nature of scattering and may not be appropriate to describe large angle scattering near the source location. As such, a full numerical treatment of scattering \citep[e.g.][]{Thejappa2008, Bian2019, Kontar2019}, in combination with interferometric imaging, is necessary to fully understand radio wave propagation in the turbulent coronal plasma. In order to definitively determine $\varepsilon$, more information on the power spectrum of density fluctuations and the scales on which radio scattering most effectively occurs is needed.

\section{Conclusion}
In summary, a Type IIIb radio burst was observed with LOFAR on 17 October 2015 at approximately 13:21:00 UTC. The bandwidth of an individual striation at 34.76~MHz suggests an FWHM source size of $3.18^{\prime\prime}$. Directly fitting visibilities to avoid effects of de-convolution algorithms reveals FWHM source sizes in the major and minor axes of $18.8^\prime$ $\pm 0.1^\prime$ and $10.2^\prime$ $\pm 0.1^\prime$, respectively. The source is located at $-1312^{\prime\prime}, -1064^{\prime\prime}$ from the solar centre. 
Having corrected for radio wave propagation in the ionosphere, we conclude that scattering from electron density fluctuations in the solar corona is the main cause of source broadening. We discuss how values for the r.m.s relative electron density fluctuations determined from numerical models and compared to tied-array observations may be overestimated. 
In the future, a combination of remote observations from LOFAR and \textit{in situ} measurements of plasma properties from the PSP and Solar Orbiter \citep{Muller2013,Muller2020} at a variety of heliocentric distances in the corona and solar wind will be needed to form a more complete picture of coronal turbulence.

\begin{acknowledgements}
We would like to thank our referee, who's insightful comments have improved this paper significantly.
P.C.M is supported by a Government of Ireland Studentship from the Irish Research Council (IRC). E.P.C is supported by the Schr\"odinger Fellowship at DIAS. A skymodel for the calibrator source was provided by the radio observatory at ASTRON. The LOFAR data used in this analysis was obtained from the Long Term Archive (LTA) under project code LC4\_001, which was part of a Solar and Space Weather Key Science Project (KSP) observing campaign. The AIA data used is courtesy of NASA/SDO and the AIA, EVE, and HMI science teams. The authors would like to thank the members of the Solar and Space Weather KSP for helpful discussions at a number of conferences and workshops. The authors would like to thank V. Krupar who provided their calculated values of $\varepsilon$ from \cite{Krupar2020}. P.C.M in particular would like to thank E. Kontar for useful discussion concerning the production of source sizes directly from visibility measurements.
Python scripts used for this analysis can be viewed at \url{https://github.com/murphp30/vis_scripts}, these made use of a number of open source python packages. Most notably; Matplotlib v3.0.2 \citep{Hunter2007}, NumPy v1.17.2 \citep{VanDerWalt2011}, Sunpy v1.1.2 \citep{Mumford2015}, Astropy v4.0.1 \citep{Robitaille2013,Price-Whelan2018}, LMFIT v0.9.14 \citep{Newville2014} and emcee v3.0.1 \citep{Foreman-Mackey2012}. The code used in this analysis was developed with the use of IPython v7.2.0 \citep{Perez2007}.
\end{acknowledgements}

\bibliographystyle{aa}
\bibliography{references}

\begin{thebibliography}{57}
\expandafter\ifx\csname natexlab\endcsname\relax\def\natexlab#1{#1}\fi

\bibitem[{Anantharamaiah {et~al.}(1994)Anantharamaiah, Gothoskar, \&
  Cornwell}]{Anantharamaiah1994}
Anantharamaiah, K.~R., Gothoskar, P., \& Cornwell, T.~J. 1994, Journal of
  Astrophysics and Astronomy, 15, 387

\bibitem[{Bastian(1994)}]{Bastian1994}
Bastian, T.~S. 1994, The Astrophysical Journal, 426, 774

\bibitem[{Bian {et~al.}(2019)Bian, Emslie, \& Kontar}]{Bian2019}
Bian, N.~H., Emslie, A.~G., \& Kontar, E.~P. 2019, The Astrophysical Journal,
  873, 33

\bibitem[{Bisoi {et~al.}(2014)Bisoi, Janardhan, Ingale, Subramanian,
  Ananthakrishnan, Tokumaru, \& Fujiki}]{Bisoi2014}
Bisoi, S.~K., Janardhan, P., Ingale, M., {et~al.} 2014, The Astrophysical
  Journal, 795, 69

\bibitem[{Briggs(1995)}]{Briggs1995}
Briggs, D.~S. 1995, American Astronomical Society, 187th AAS Meeting,
  id.112.02; Bulletin of the American Astronomical Society, Vol. 27, p.1444,
  187, 112.02

\bibitem[{Chrysaphi {et~al.}(2018)Chrysaphi, Kontar, Holman, \&
  Temmer}]{Chrysaphi2018}
Chrysaphi, N., Kontar, E.~P., Holman, G.~D., \& Temmer, M. 2018, The
  Astrophysical Journal, 868, 79

\bibitem[{Cornwell(2008)}]{Cornwell2008}
Cornwell, T.~J. 2008, IEEE Journal on Selected Topics in Signal Processing, 2,
  793

\bibitem[{de~Gasperin {et~al.}(2019)de~Gasperin, Dijkema, Drabent, Mevius,
  Rafferty, van Weeren, Br{\"{u}}ggen, Callingham, Emig, Heald, Intema,
  Morabito, Offringa, Oonk, Orr{\`{u}}, R{\"{o}}ttgering, Sabater, Shimwell,
  Shulevski, \& Williams}]{DeGasperin2019}
de~Gasperin, F., Dijkema, T.~J., Drabent, A., {et~al.} 2019, Astronomy {\&}
  Astrophysics, 622, A5

\bibitem[{de~La~Noe(1975)}]{DeLaNoe1975}
de~La~Noe, J. 1975, Astronomy and Astrophysics, 43, 201

\bibitem[{de~La~Noe \& Boischot(1972)}]{DeLaNoe1972}
de~La~Noe, J. \& Boischot, A. 1972, Astronomy and Astrophysics, 20, 55

\bibitem[{Dewdney {et~al.}(2009)Dewdney, Hall, Schilizzi, \&
  Lazio}]{Dewdney2009}
Dewdney, P.~E., Hall, P.~J., Schilizzi, R.~T., \& Lazio, T. J.~L. 2009,
  Proceedings of the IEEE, 97, 1482

\bibitem[{Ellis(1969)}]{Ellis1969}
Ellis, G. 1969, Australian Journal of Physics, 22, 177

\bibitem[{Ellis \& McCulloch(1967)}]{Ellis1967}
Ellis, G. \& McCulloch, P. 1967, Australian Journal of Physics, 20, 583

\bibitem[{Fokker(1965)}]{Fokker1965}
Fokker, A.~D. 1965, Bulletin of the Astronomical Institutes of the Netherlands,
  18, 111

\bibitem[{Foreman-Mackey {et~al.}(2012)Foreman-Mackey, Hogg, Lang, \&
  Goodman}]{Foreman-Mackey2012}
Foreman-Mackey, D., Hogg, D.~W., Lang, D., \& Goodman, J. 2012

\bibitem[{Fox {et~al.}(2016)Fox, Velli, Bale, Decker, Driesman, Howard, Kasper,
  Kinnison, Kusterer, Lario, Lockwood, McComas, Raouafi, \& Szabo}]{Fox2016}
Fox, N.~J., Velli, M.~C., Bale, S.~D., {et~al.} 2016, Space Science Reviews,
  204, 7

\bibitem[{Gordovskyy {et~al.}(2019)Gordovskyy, Kontar, Browning, \&
  Kuznetsov}]{Gordovskyy2019}
Gordovskyy, M., Kontar, E., Browning, P., \& Kuznetsov, A. 2019, The
  Astrophysical Journal, 873, 48

\bibitem[{H{\"{o}}gbom(1974)}]{Hogbom1974}
H{\"{o}}gbom, J.~A. 1974, Astronomy {\&} Astrophysics Supplemental, 15, 417

\bibitem[{Hunter(2007)}]{Hunter2007}
Hunter, J.~D. 2007, Computing in Science and Engineering, 9, 99

\bibitem[{Hurford {et~al.}(2002)Hurford, Schmahl, Schwartz, Conway, Aschwanden,
  Csillaghy, Dennis, Johns-Krull, Krucker, Lin, McTiernan, Metcalf, Sato, \&
  Smith}]{Hurford2002}
Hurford, G.~J., Schmahl, E.~J., Schwartz, R.~A., {et~al.} 2002, Solar Physics,
  210, 61

\bibitem[{Ingale {et~al.}(2015)Ingale, Subramanian, \& Cairns}]{Ingale2015}
Ingale, M., Subramanian, P., \& Cairns, I. 2015, Monthly Notices of the Royal
  Astronomical Society, 447, 3486

\bibitem[{Kontar {et~al.}(2019)Kontar, Chen, Chrysaphi, Jeffrey, Emslie,
  Krupar, Maksimovic, Gordovskyy, \& Browning}]{Kontar2019}
Kontar, E.~P., Chen, X., Chrysaphi, N., {et~al.} 2019, The Astrophysical
  Journal, 884, 122

\bibitem[{Kontar {et~al.}(2010)Kontar, Hannah, Jeffrey, \&
  Battaglia}]{Kontar2010}
Kontar, E.~P., Hannah, I.~G., Jeffrey, N. L.~S., \& Battaglia, M. 2010, The
  Astrophysical Journal, 717, 250

\bibitem[{Kontar {et~al.}(2017)Kontar, Yu, Kuznetsov, Emslie, Alcock, Jeffrey,
  Melnik, Bian, \& Subramanian}]{Kontar2017}
Kontar, E.~P., Yu, S., Kuznetsov, A.~A., {et~al.} 2017, Nature Communications,
  8, 1515

\bibitem[{Krupar {et~al.}(2018)Krupar, Maksimovic, Kontar, Zaslavsky, Santolik,
  Soucek, Kruparova, Eastwood, \& Szabo}]{Krupar2018}
Krupar, V., Maksimovic, M., Kontar, E.~P., {et~al.} 2018, The Astrophysical
  Journal, 857, 82

\bibitem[{Krupar {et~al.}(2020)Krupar, Szabo, Maksimovic, Kruparova, Kontar,
  Balmaceda, Bonnin, Bale, Pulupa, Malaspina, Bonnell, Harvey, Goetz, de~Wit,
  MacDowall, Kasper, Case, Korreck, Larson, Livi, Stevens, Whittlesey, \&
  Hegedus}]{Krupar2020}
Krupar, V., Szabo, A., Maksimovic, M., {et~al.} 2020

\bibitem[{Lemen {et~al.}(2012)Lemen, Title, Akin, Boerner, Chou, Drake, Duncan,
  Edwards, Friedlaender, Heyman, Hurlburt, Katz, Kushner, Levay, Lindgren,
  Mathur, McFeaters, Mitchell, Rehse, Schrijver, Springer, Stern, Tarbell,
  Wuelser, Wolfson, Yanari, Bookbinder, Cheimets, Caldwell, Deluca, Gates,
  Golub, Park, Podgorski, Bush, Scherrer, Gummin, Smith, Auker, Jerram, Pool,
  Soufli, Windt, Beardsley, Clapp, Lang, \& Waltham}]{Lemen2012}
Lemen, J.~R., Title, A.~M., Akin, D.~J., {et~al.} 2012, Solar Physics, 275, 17

\bibitem[{Lonsdale {et~al.}(2009)Lonsdale, Cappallo, Morales, Briggs,
  Benkevitch, Bowman, Bunton, Burns, Corey, DeSouza, Doeleman, Derome,
  Deshpande, Gopala, Greenhill, Herne, Hewitt, Kamini, Kasper, Kincaid, Kocz,
  Kowald, Kratzenberg, Kumar, Lynch, Madhavi, Matejek, Mitchell, Morgan,
  Oberoi, Ord, Pathikulangara, Prabu, Rogers, Roshi, Salah, Sault, Shankar,
  Srivani, Stevens, Tingay, Vaccarella, Waterson, Wayth, Webster, Whitney,
  Williams, \& Williams}]{Lonsdale2009}
Lonsdale, C.~J., Cappallo, R.~J., Morales, M.~F., {et~al.} 2009, Proceedings of
  the IEEE, 97, 1497

\bibitem[{McLean \& Labrum(1985)}]{McLean1985}
McLean, D.~J. \& Labrum, N.~R. 1985, Cambridge and New York, Cambridge
  University Press, 1985, 527 p. For individual items see A87-13852 to
  A87-13867.

\bibitem[{Melnik {et~al.}(2010)Melnik, Rucker, Konovalenko, Shevchuk, Abranin,
  Dorovskyy, \& Lecacheux}]{Melnik2010b}
Melnik, V.~N., Rucker, H.~O., Konovalenko, A.~A., {et~al.} 2010, in AIP
  Conference Proceedings, Vol. 1206, 445--449

\bibitem[{Morosan {et~al.}(2019)Morosan, Carley, Hayes, Murray, Zucca, Fallows,
  McCauley, Kilpua, Mann, Vocks, \& Gallagher}]{Morosan2019b}
Morosan, D.~E., Carley, E.~P., Hayes, L.~A., {et~al.} 2019, Nature Astronomy,
  3, 452

\bibitem[{Morosan {et~al.}(2014)Morosan, Gallagher, Zucca, Fallows, Carley,
  Mann, Bisi, Kerdraon, Konovalenko, MacKinnon, Rucker, Thid{\'{e}},
  Magdaleni{\'{c}}, Vocks, Reid, Anderson, Asgekar, Avruch, Bentum, Bernardi,
  Best, Bonafede, Bregman, Breitling, Broderick, Br{\"{u}}ggen, Butcher,
  Ciardi, Conway, de~Gasperin, de~Geus, Deller, Duscha, Eisl{\"{o}}ffel,
  Engels, Falcke, Ferrari, Frieswijk, Garrett, Grie{\ss}meier, Gunst, Hassall,
  Hessels, Hoeft, H{\"{o}}randel, Horneffer, Iacobelli, Juette, Karastergiou,
  Kondratiev, Kramer, Kuniyoshi, Kuper, Maat, Markoff, McKean, Mulcahy, Munk,
  Nelles, Norden, Orru, Paas, Pandey-Pommier, Pandey, Pietka, Pizzo, Polatidis,
  Reich, R{\"{o}}ttgering, Scaife, Schwarz, Serylak, Smirnov, Stappers,
  Stewart, Tagger, Tang, Tasse, Thoudam, Toribio, Vermeulen, van Weeren,
  Wucknitz, Yatawatta, \& Zarka}]{Morosan2014}
Morosan, D.~E., Gallagher, P.~T., Zucca, P., {et~al.} 2014, Astronomy {\&}
  Astrophysics, 568, A67

\bibitem[{M{\"{u}}ller {et~al.}(2020)M{\"{u}}ller, Cyr, Zouganelis, Gilbert,
  Marsden, Nieves-Chinchilla, Antonucci, Auch{\`{e}}re, Berghmans, Horbury,
  Howard, Krucker, Maksimovic, Owen, Rochus, Rodriguez-Pacheco, Romoli,
  Solanki, Bruno, Carlsson, Fludra, Harra, Hassler, Livi, Louarn, Peter,
  Sch{\"{u}}hle, Teriaca, Iniesta, Wimmer-Schweingruber, Marsch, Velli,
  De~Groof, Walsh, \& Williams}]{Muller2020}
M{\"{u}}ller, D., Cyr, O. C.~S., Zouganelis, I., {et~al.} 2020, Astronomy {\&}
  Astrophysics, 642, A1

\bibitem[{M{\"{u}}ller {et~al.}(2013)M{\"{u}}ller, Marsden, St.~Cyr, \&
  Gilbert}]{Muller2013}
M{\"{u}}ller, D., Marsden, R.~G., St.~Cyr, O.~C., \& Gilbert, H.~R. 2013, Solar
  Physics, 285, 25

\bibitem[{Mumford {et~al.}(2015)Mumford, Christe, P{\'{e}}rez-Su{\'{a}}rez,
  Ireland, Shih, Inglis, Liedtke, Hewett, Mayer, Hughitt, Freij, Meszaros,
  Bennett, Malocha, Evans, Agrawal, Leonard, Robitaille, Mampaey, Campos-Rozo,
  \& Kirk}]{Mumford2015}
Mumford, S.~J., Christe, S., P{\'{e}}rez-Su{\'{a}}rez, D., {et~al.} 2015,
  Computational Science and Discovery, 8

\bibitem[{Newville {et~al.}(2014)Newville, Stensitzki, Allen, \&
  Ingargiola}]{Newville2014}
Newville, M., Stensitzki, T., Allen, D.~B., \& Ingargiola, A. 2014

\bibitem[{Offringa {et~al.}(2014)Offringa, McKinley, Hurley-Walker, Briggs,
  Wayth, Kaplan, Bell, Feng, Neben, Hughes, Rhee, Murphy, Bhat, Bernardi,
  Bowman, Cappallo, Corey, Deshpande, Emrich, Ewall-Wice, Gaensler, Goeke,
  Greenhill, Hazelton, Hindson, Johnston-Hollitt, Jacobs, Kasper, Kratzenberg,
  Lenc, Lonsdale, Lynch, McWhirter, Mitchell, Morales, Morgan, Kudryavtseva,
  Oberoi, Ord, Pindor, Procopio, Prabu, Riding, Roshi, Shankar, Srivani,
  Subrahmanyan, Tingay, Waterson, Webster, Whitney, Williams, \&
  Williams}]{Offringa2014}
Offringa, A.~R., McKinley, B., Hurley-Walker, N., {et~al.} 2014, Monthly
  Notices of the Royal Astronomical Society, 444, 606

\bibitem[{Offringa \& Smirnov(2017)}]{Offringa2017}
Offringa, A.~R. \& Smirnov, O. 2017

\bibitem[{P{\'{e}}rez \& Granger(2007)}]{Perez2007}
P{\'{e}}rez, F. \& Granger, B.~E. 2007, Computing in Science and Engineering,
  9, 21

\bibitem[{Price-Whelan {et~al.}(2018)Price-Whelan, Sip{\H{o}}cz, G{\"{u}}nther,
  Lim, Crawford, Conseil, Shupe, Craig, Dencheva, Ginsburg, VanderPlas,
  Bradley, P{\'{e}}rez-Su{\'{a}}rez, de~Val-Borro, Aldcroft, Cruz, Robitaille,
  Tollerud, Ardelean, Babej, Bach, Bachetti, Bakanov, Bamford, Barentsen,
  Barmby, Baumbach, Berry, Biscani, Boquien, Bostroem, Bouma, Brammer, Bray,
  Breytenbach, Buddelmeijer, Burke, Calderone, Rodr{\'{i}}guez, Cara, Cardoso,
  Cheedella, Copin, Corrales, Crichton, D’Avella, Deil, Depagne, Dietrich,
  Donath, Droettboom, Earl, Erben, Fabbro, Ferreira, Finethy, Fox, Garrison,
  Gibbons, Goldstein, Gommers, Greco, Greenfield, Groener, Grollier, Hagen,
  Hirst, Homeier, Horton, Hosseinzadeh, Hu, Hunkeler, Ivezi{\'{c}}, Jain,
  Jenness, Kanarek, Kendrew, Kern, Kerzendorf, Khvalko, King, Kirkby, Kulkarni,
  Kumar, Lee, Lenz, Littlefair, Ma, Macleod, Mastropietro, McCully, Montagnac,
  Morris, Mueller, Mumford, Muna, Murphy, Nelson, Nguyen, Ninan, N{\"{o}}the,
  Ogaz, Oh, Parejko, Parley, Pascual, Patil, Patil, Plunkett, Prochaska,
  Rastogi, Janga, Sabater, Sakurikar, Seifert, Sherbert, Sherwood-Taylor, Shih,
  Sick, Silbiger, Singanamalla, Singer, Sladen, Sooley, Sornarajah, Streicher,
  Teuben, Thomas, Tremblay, Turner, Terr{\'{o}}n, Kerkwijk, de~la Vega,
  Watkins, Weaver, Whitmore, Woillez, \& Zabalza}]{Price-Whelan2018}
Price-Whelan, A.~M., Sip{\H{o}}cz, B.~M., G{\"{u}}nther, H.~M., {et~al.} 2018,
  The Astronomical Journal, 156, 123

\bibitem[{Reid \& Kontar(2017)}]{Reid2017}
Reid, H.~A. \& Kontar, E.~P. 2017, Astronomy and Astrophysics, 606

\bibitem[{Reid \& Kontar(2010)}]{Reid2010}
Reid, H. A.~S. \& Kontar, E.~P. 2010, The Astrophysical Journal, 721, 864

\bibitem[{Reid \& Ratcliffe(2014)}]{Reid2014}
Reid, H. A.~S. \& Ratcliffe, H. 2014, Research in Astronomy and Astrophysics,
  14, 773

\bibitem[{Riddle(1974)}]{Riddle1974}
Riddle, A.~C. 1974, Solar Physics, 35, 153

\bibitem[{Robitaille {et~al.}(2013)Robitaille, Tollerud, Greenfield,
  Droettboom, Bray, Aldcroft, Davis, Ginsburg, Price-Whelan, Kerzendorf,
  Conley, Crighton, Barbary, Muna, Ferguson, Grollier, Parikh, Nair,
  G{\"{u}}nther, Deil, Woillez, Conseil, Kramer, Turner, Singer, Fox, Weaver,
  Zabalza, Edwards, Azalee~Bostroem, Burke, Casey, Crawford, Dencheva, Ely,
  Jenness, Labrie, Lim, Pierfederici, Pontzen, Ptak, Refsdal, Servillat, \&
  Streicher}]{Robitaille2013}
Robitaille, T.~P., Tollerud, E.~J., Greenfield, P., {et~al.} 2013, Astronomy
  and Astrophysics, 558

\bibitem[{Sasikumar~Raja {et~al.}(2016)Sasikumar~Raja, Ingale, Ramesh,
  Subramanian, Manoharan, \& Janardhan}]{SasikumarRaja2016}
Sasikumar~Raja, K., Ingale, M., Ramesh, R., {et~al.} 2016

\bibitem[{Sharma \& Oberoi(2020)}]{Sharma2020}
Sharma, R. \& Oberoi, D. 2020

\bibitem[{Sharykin {et~al.}(2018)Sharykin, Kontar, \& Kuznetsov}]{Sharykin2018}
Sharykin, I.~N., Kontar, E.~P., \& Kuznetsov, A.~A. 2018, Solar Physics, 293,
  115

\bibitem[{Steinberg {et~al.}(1971)Steinberg, Aubier-Giraud, Leblanc, \&
  Boischot}]{Steinberg1971}
Steinberg, J.~L., Aubier-Giraud, M., Leblanc, Y., \& Boischot, A. 1971,
  Astronomy and Astrophysics, 10, 362

\bibitem[{Stewart(1972)}]{Stewart1972}
Stewart, R.~T. 1972, Publications of the Astronomical Society of Australia, 2,
  100

\bibitem[{Takakura \& Yousef(1975)}]{Takakura1975}
Takakura, T. \& Yousef, S. 1975, Solar Physics, 40, 421

\bibitem[{Thejappa \& MacDowall(2008)}]{Thejappa2008}
Thejappa, G. \& MacDowall, R.~J. 2008, The Astrophysical Journal, 676, 1338

\bibitem[{Thejappa {et~al.}(2007)Thejappa, MacDowall, \& Kaiser}]{Thejappa2007}
Thejappa, G., MacDowall, R.~J., \& Kaiser, M.~L. 2007, The Astrophysical
  Journal, 671, 894

\bibitem[{Van Der~Walt {et~al.}(2011)Van Der~Walt, Colbert, \&
  Varoquaux}]{VanDerWalt2011}
Van Der~Walt, S., Colbert, S.~C., \& Varoquaux, G. 2011, Computing in Science
  and Engineering, 13, 22

\bibitem[{van Diepen {et~al.}(2018)van Diepen, Dijkema, \&
  Offringa}]{VanDiepen2018}
van Diepen, G., Dijkema, T.~J., \& Offringa, A. 2018, ascl, ascl:1804.003

\bibitem[{van Haarlem {et~al.}(2013)van Haarlem, Wise, Gunst, Heald, McKean,
  Hessels, de~Bruyn, Nijboer, Swinbank, Fallows, Brentjens, Nelles, Beck,
  Falcke, Fender, H{\"{o}}randel, Koopmans, Mann, Miley, R{\"{o}}ttgering,
  Stappers, Wijers, Zaroubi, Akker, Alexov, Anderson, Anderson, van Ardenne,
  Arts, Asgekar, Avruch, Batejat, B{\"{a}}hren, Bell, Bell, van Bemmel,
  Bennema, Bentum, Bernardi, Best, B{\^{i}}rzan, Bonafede, Boonstra, Braun,
  Bregman, Breitling, van~de Brink, Broderick, Broekema, Brouw, Br{\"{u}}ggen,
  Butcher, van Cappellen, Ciardi, Coenen, Conway, Coolen, Corstanje, Damstra,
  Davies, Deller, Dettmar, van Diepen, Dijkstra, Donker, Doorduin, Dromer,
  Drost, van Duin, Eisl{\"{o}}ffel, van Enst, Ferrari, Frieswijk, Gankema,
  Garrett, de~Gasperin, Gerbers, de~Geus, Grie{\ss}meier, Grit, Gruppen,
  Hamaker, Hassall, Hoeft, Holties, Horneffer, van~der Horst, van Houwelingen,
  Huijgen, Iacobelli, Intema, Jackson, Jelic, de~Jong, Juette, Kant,
  Karastergiou, Koers, Kollen, Kondratiev, Kooistra, Koopman, Koster,
  Kuniyoshi, Kramer, Kuper, Lambropoulos, Law, van Leeuwen, Lemaitre, Loose,
  Maat, Macario, Markoff, Masters, McKay-Bukowski, Meijering, Meulman, Mevius,
  Middelberg, Millenaar, Miller-Jones, Mohan, Mol, Morawietz, Morganti,
  Mulcahy, Mulder, Munk, Nieuwenhuis, van Nieuwpoort, Noordam, Norden, Noutsos,
  Offringa, Olofsson, Omar, Orr{\'{u}}, Overeem, Paas, Pandey-Pommier, Pandey,
  Pizzo, Polatidis, Rafferty, Rawlings, Reich, de~Reijer, Reitsma, Renting,
  Riemers, Rol, Romein, Roosjen, Ruiter, Scaife, van~der Schaaf, Scheers,
  Schellart, Schoenmakers, Schoonderbeek, Serylak, Shulevski, Sluman, Smirnov,
  Sobey, Spreeuw, Steinmetz, Sterks, Stiepel, Stuurwold, Tagger, Tang, Tasse,
  Thomas, Thoudam, Toribio, van~der Tol, Usov, van Veelen, van~der Veen, ter
  Veen, Verbiest, Vermeulen, Vermaas, Vocks, Vogt, de~Vos, van~der Wal, van
  Weeren, Weggemans, Weltevrede, White, Wijnholds, Wilhelmsson, Wucknitz,
  Yatawatta, Zarka, Zensus, van Zwieten, \& van Zwieten}]{VanHaarlem2013b}
van Haarlem, M.~P., Wise, M.~W., Gunst, A.~W., {et~al.} 2013, Astronomy {\&}
  Astrophysics, Volume 556, id.A2, 53 pp., 556

\bibitem[{Zucca {et~al.}(2018)Zucca, Morosan, Rouillard, Fallows, Gallagher,
  Magdalenic, Klein, Mann, Vocks, Carley, Bisi, Kontar, Rothkaehl, Dabrowski,
  Krankowski, Anderson, Asgekar, Bell, Bentum, Best, Blaauw, Breitling,
  Broderick, Brouw, Br{\"{u}}ggen, Butcher, Ciardi, Geus, Deller, Duscha,
  Eisl{\"{o}}ffel, Garrett, Grie{\ss}meier, Gunst, Heald, Hoeft,
  H{\"{o}}randel, Iacobelli, Juette, Karastergiou, Leeuwen, McKay-Bukowski,
  Mulder, Munk, Nelles, Orru, Paas, Pandey, Pekal, Pizzo, Polatidis, Reich,
  Rowlinson, Schwarz, Shulevski, Sluman, Smirnov, Sobey, Soida, Thoudam,
  Toribio, Vermeulen, van Weeren, Wucknitz, \& Zarka}]{Zucca2018}
Zucca, P., Morosan, D.~E., Rouillard, A.~P., {et~al.} 2018, Astronomy {\&}
  Astrophysics, 615, A89

\end{thebibliography}

\end{document}